\documentclass[sigconf]{acmart}

\AtBeginDocument{%
  }

\usepackage{array}
\usepackage{multirow}
\begin{document}

\title{Strategic Optimization and Challenges of Large Language Models in Object-Oriented Programming}

\author{Zinan Wang}
\email{zinanwang-paper@outlook.com}









\begin{abstract}
In the area of code generation research\cite{song2019survey}, the emphasis has transitioned from crafting individual functions to developing class-level method code that integrates contextual information. This shift has brought several benchmarks such as ClassEval\cite{du2023classeval} and CoderEval\cite{yu2024codereval}, which consider class-level contexts. Nevertheless, the influence of specific contextual factors at the method level remains less explored. These factors include the method's own description, its interactions with other modules, and the broader project context. The varied impacts of these factors on code generation outcomes—such as pass rates, error type distributions, and developer support—could also have economic implications, notably in terms of the token consumption required for API calls. Furthermore, the extent of a method's external module interactions, or ''coupling'', significantly influences code generation, yet quantitative analyses of this impact are sparse.

Our research focused on method-level code generation within the Object-Oriented Programming (OOP) framework. We devised experiments that varied the extent of contextual information in the prompts, ranging from method-specific to project-wide details. We introduced the innovative metric of "Prompt-Token Cost-Effectiveness" to evaluate the economic viability of incorporating additional contextual layers. Our findings indicate that prompts enriched with method invocation details yield the highest cost-effectiveness. Additionally, our study revealed disparities among Large Language Models (LLMs) regarding error type distributions and the level of assistance they provide to developers. Notably, larger LLMs do not invariably perform better. We also observed that tasks with higher degrees of coupling present more substantial challenges, suggesting that the choice of LLM should be tailored to the task's coupling degree. For example, GPT-4 exhibited improved performance in low-coupling scenarios, whereas GPT-3.5 seemed better suited for tasks with high coupling. By meticulously curating prompt content and selecting the appropriate LLM, developers can optimize code quality while maximizing cost-efficiency during the development process.
\end{abstract}





\maketitle

\section{Introduction}
\label{sec:intro}
In the past, automated code generation primarily relied on simple templates or rule engines\cite{zhang2023survey}. While effective in certain scenarios, these methods often lacked flexibility and the ability to adapt to complex requirements\cite{dehaerne2022code}. With the rapid advancement of Natural Language Processing(NLP)\cite{vaswani2017attention}, LLMs, particularly OpenAI's GPT series, have made significant strides in the field of intelligent code generation\cite{ouyang2023llm}. These models not only comprehend and generate human language but have also extended their capabilities to understand complex programming languages and autonomously generate code. From automatically completing code snippets to generating entire functional code structures, the application of LLMs is driving advancements in software development.

Although LLMs have made progress in code generation, existing research has primarily focused on simple isolated code tasks, such as generating solutions for algorithmic problems, with little consideration for tasks with complex contextual information in practical scenarios. These studies and benchmarks, such as HumanEval, often concentrate on evaluating the model's ability to generate single functions or small code blocks. There is a lack of consideration for scenarios involving complex contexts in actual development environments. Regarding existing OOP code generation test frameworks, like ClassEval, while they consider class-level context such as basic class structures and frameworks, they do not deeply analyze the impact of varying degrees of contextual information on the generated results. These contextual details can range from specific descriptions of method functionalities to encompassing other modules invoked by them, to broader project integration information descriptions.

Besides, considering that tasks with higher coupling are more complex, this increases the difficulty of generation. Additionally, the degree of coupling itself is a characteristic of the target method. To determine which LLM is more suitable for specific tasks, it is essential to study the degree of coupling. Frameworks like CoderEval have begun to focus on code independence (i.e., different levels of independence ranging from "completely standalone" to "requiring other modules in the project to run"). However, there is still a lack of systematic quantitative analysis on the coupling degree of the target code. This involves factors such as the number of APIs, variables, and classes called by the target method, and how this coupling affects the quality of the generated code.

To address these research gaps, this experiment was designed to control the prompt richness using prompt levels (i.e., from \textbf{Base-Level} containing only the name and signature of the target method to \textbf{Project-Level} containing information related to the entire project). Based on this, the study aimed to analyze in depth the specific impact of different prompt richness levels on the quality of code generation. Firstly, we measured the variation of Pass@K under different prompt levels. Then, we analyzed the economic aspect of the additional prompt information for each level based on the designed "Prompt-Token Cost-Effectiveness." Additionally, we conducted an analysis of the distribution of errors in the generated code (such as syntax errors and hallucination errors\cite{huang2023survey}) using a combination of code analysis tools and expert evaluation. To measure the extent to which the generated code assists developers, we introduced the metric "Helpfulness-Value"\cite{evtikhiev2023out} and analyzed its distribution based on expert assessments. Regarding the coupling degree of target methods, we quantitatively described the degree of coupling based on the concept of "Fan-Out."\cite{murgia2012refactoring} Subsequently, we analyzed its impact on Pass@K of the generated code.

The research findings indicate that prompts containing method invocation information perform best in terms of improving cost-effectiveness. Additionally, as the amount of prompt information increases, although the overall number of errors decreases, there is a rising trend in specific types of errors (such as overdesign\cite{huang2023survey}). Furthermore, there are significant differences in the distribution of the helpfulness value of generated code among LLMs. The experiments also suggest that when selecting an LLM, the coupling degree of the target method should be considered. For instance, GPT-4 performs better in low-coupling tasks, while GPT-3.5 may be more suitable for medium to high-coupling tasks.These findings offer practical guidance for developers, aiding them in optimizing their code generation strategies.

The content of this paper is organized as follows: Section \ref{sec:bg} discusses the reasons and technical foundations for the experiment. Section \ref{sec:rl} describes the progress of research relevant to the experiment in this paper. Section \ref{sec:sd} outlines the overall design of the study, including the research framework, research settings, metrics used, and research questions. Section \ref{sec:sr} elaborates on and discusses the results of the research questions. The \ref{sec:dis} section discusses the threats to this paper and compiles suggestions for developers based on the experimental results. The \ref{sec:conc} section provides a conclusion of the entire study.

\section{Background}
\label{sec:bg}

Current research exhibits a notable deficiency in attention and investigation into method-level code generation for OOP. In this paper, "method-level code generation" is defined as the process of generating code for a specific method of a given object within an OOP project, utilizing Large Language Models (LLMs) based on contextual information and precise method descriptions. This study selects the CoderEval\cite{yu2024codereval} dataset as the cornerstone of its experiments, aggregating 460 self-contained executable code generation tasks from genuine open-source projects, encompassing both Python and Java languages. The rationale behind choosing this dataset is to offer an evaluation platform that closely resembles a real development environment, aiming to more accurately capture and exhibit the performance and challenges of automated code generation tools in practical application scenarios.

Specifically, we focus on the Java subset of the CoderEval dataset, comprising 230 tasks, which we have significantly refined and applied to ensure that our experimental design and data usage reflect the real-world demands of Java development as comprehensively as possible, particularly in terms of code quality assessment and the practical value of generation tasks. To augment the practicality and efficiency of the experiments, we have meticulously dissected and reassembled the CoderEval dataset. We have segmented and adjusted the parts of the code description, achieving a precise gradation of code hints. This step has established a solid foundation for subsequent data processing and analysis.

\begin{figure*}[h]
  \centering
  \includegraphics[width=\textwidth]{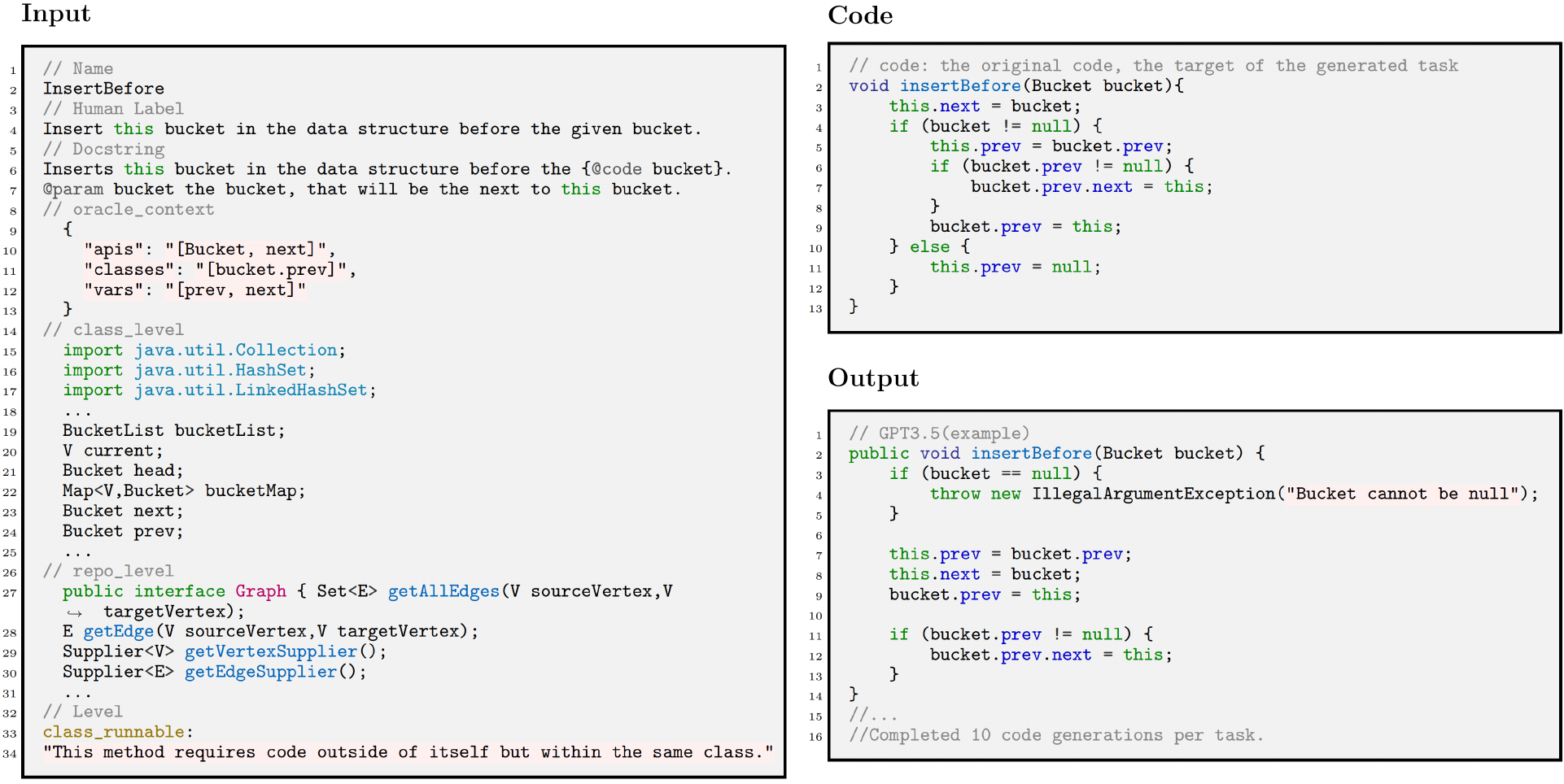}
  \caption{Code Generate Example: Input: Prompt information for LLMs. Code: Reference code for the target method. Output: Code generated by LLMs (example).}
  \label{fig:CoderEval}
\end{figure*}

As depicted in the Figure \ref{fig:CoderEval}, we present a code generation task at the highest level of hint precision. The definitions of precision levels and hint content will be elaborated on in the following sections. Generally, "input" signifies the information provided, which is transformed into hints for the generation task; "code" denotes the target code to be generated; "output" illustrates an example of the generated code. For each task, the experiment conducts ten iterations of code generation to ensure the stability and reliability of the outcomes.

The term "input" encompasses exhaustive information from the fundamental definition of methods to the comprehensive structure of the entire project. It includes the method's \textbf{signature} and \textbf{name}, which provide the model with basic invocation and identification mechanisms. A succinct description (\textbf{human\_label}) and an elaborate \textbf{docstring} offer detailed explanations that aid the model in understanding the method's purpose and operational specifics. The \textbf{oracle\_context} supplies additional contextual information, such as utilized APIs and relevant variables, enhancing the model's adaptability to the code environment. The \textbf{class\_level} and \textbf{repo\_level} broaden the scope to detailed structures at the class and project levels, showcasing the internal makeup of classes and the extensive logical relationships within the project. Lastly, the \textbf{level} section indicates the code snippet's self-containment, denoting its independence within the overall project. This amalgamation of information ensures the generated code's precision and its high coordination with the project's overarching architecture and functional requirements.

\section{Related Works}
\label{sec:rl}

\subsection{LLMs and Code Generation}
With the emergence of LLMs, significant achievements have been made in the field of NLP \cite{liu2023summary,wang2023gpt,rubel2020biobertpt,singhal2023towards,lazaridou2022internet}, and they are gradually being applied to various aspects of software engineering, especially code generation. These models, through pre-training and fine-tuning, can learn and generate logically coherent code snippets. For instance, Baptiste Rozière et al. propose Code Llama \cite{roziere2023code}, and Mark Chen et al. propose CodeX \cite{chen2021evaluating}, which have achieved certain results by fine-tuning for code generation tasks. They demonstrate advantages in tasks such as automated code completion, bug fixing, and direct code generation from descriptions.

Building upon these, researchers are exploring various ways to further leverage the potential of the LLMs themselves \cite{wei2022chain,hinton2015distilling,jacob2018quantization,giray2023prompt,white2023prompt,zhou2022large,sorensen2022information}. This includes proposing more advanced prompting engineering methods, such as Chain-of-Thought \cite{wei2022chain}, as proposed by Jason Wei et al., which significantly enhance the accuracy and interpretability of LLMs by introducing a process of incremental reasoning into the LLMs' inputs. Additionally, in order to reduce usage costs, researchers may employ classic resource-saving methods such as model distillation \cite{hinton2015distilling}, as proposed by Geoffrey Hinton et al., and quantization compression \cite{jacob2018quantization,gholami2022survey}. However, this trade-off can lead to performance degradation and a decrease in generalization capability.


To better evaluate the abilities of various LLMs in the code generation scenario, researchers have proposed a series of generation tasks. For instance, Mark Chen et al. propose HumanEval \cite{chen2021evaluating}, which involves having LLMs write specific programming problems and unit tests to assess whether the generated code can correctly perform the required tasks. Additionally, Shiqi Wang et al. propose ReCode \cite{wang2022recode}, which offers a suite of robustness tests for code generation models. However, most of these studies primarily focus on code generation tasks that do not require considering context.

\subsection{Code Generation in OOP}
As research deepens and LLMs continue to evolve, an increasing number of researchers are also exploring class-level and method-level code generation, such as generating method-level code for OOP\cite{fan2023large,cipriano2023gpt}. In these specific scenarios, existing evaluation metrics gradually reveal limitations, as they fail to consider the intricate contextual nuances in real-world business scenarios. Therefore, new generation tasks tailored to these specific scenarios are being suggested. Yiyang Hao et al. propose AixBench \cite{hao2022aixbench}, which focuses on method-level code generation and covers various programming problems. Hao Yu et al. propose CoderEval \cite{yu2024codereval}, designed for real coding scenarios to assess LLMs' handling of code context dependencies. Xueying Du et al. propose ClassEval \cite{du2023classeval}, which emphasizes class-level code generation, evaluating LLMs' ability to generate entire classes.

Although current evaluations and research tasks have partially addressed the gaps in the subclass of OOP method-level generation, they still fail to fully encompass the intricate aspects of the code generation process (e.g., how to select appropriate prompt content and suitable models). This research gap limits developers' flexibility in selecting and applying models in real-world development scenarios. While some new evaluation metrics, such as CrossCodeEval proposed by Yangruibo Ding et al. \cite{ding2024crosscodeeval}, have begun to consider cross-file-level contextual prompts, there is a lack of analysis on how the prompt content itself impacts the code generation results.

Furthermore, from a practical application perspective, the inference cost of LLMs is a key consideration, whether through API calls or local inference. For the same LLM, the usage cost can be directly linked to token consumption. Achieving the optimal cost-effectiveness implies minimizing token consumption while ensuring that the generated code maintains high quality. Although existing evaluation metrics and generation strategies aim to measure the model's performance ceiling, the actual cost required to approach this performance ceiling is significant. Current research mostly focuses on technical means such as quantization compression or employing Mixed Expert Models (MoE) to reduce model runtime costs, but there is relatively less research on optimizing cost-effectiveness from the user's perspective. Therefore, finding the optimal balance between cost and effectiveness from an economic standpoint is crucial for promoting the application of large models in industrial practice.

\section{Study Design}
\label{sec:sd}
\begin{figure*}[h]
  \centering
  \includegraphics[width=\textwidth]{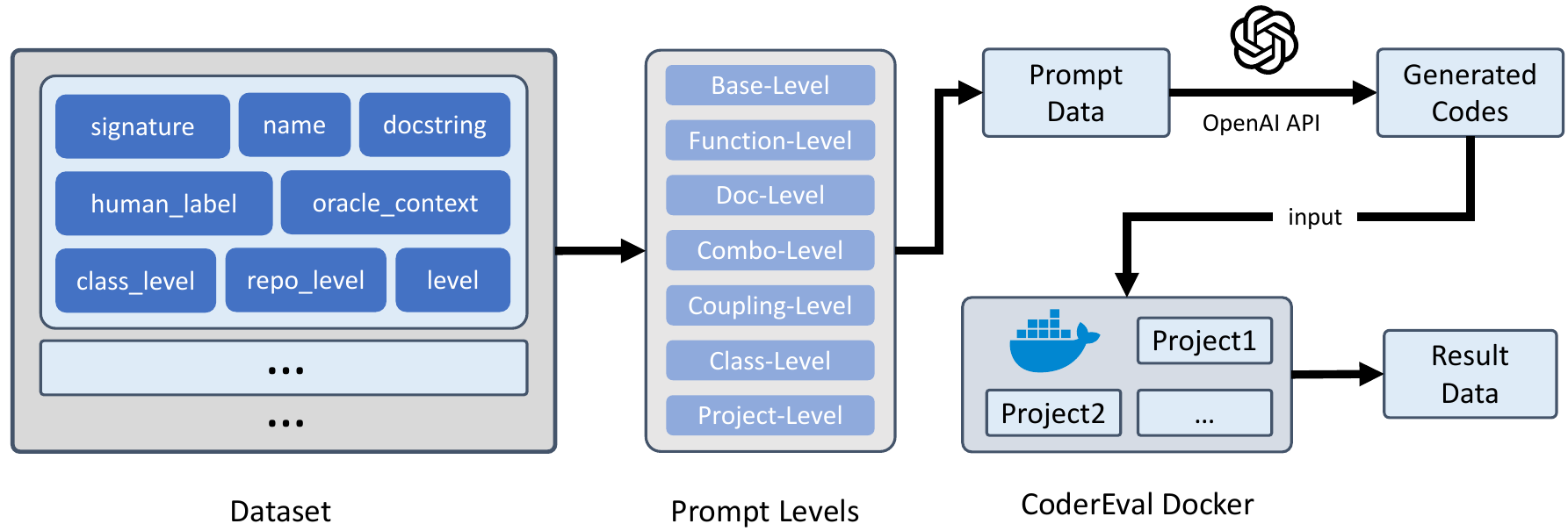}
  \caption{Research FrameWork. Following the arrows: Dataset: Contains input data describing target methods. Prompt Level: Defines varying levels of prompt word content. Prompt Data: Input information for generating code, to be tested within the CoderEval Docker environment. Generated Codes: The resulting code output. CoderEval Docker: An image containing numerous projects, utilized for assessing the quality of code generation.}
  \label{fig:FrameWork}
\end{figure*}

\subsection{Dataset}
The experiments were conducted based on the data provided by CoderEval, which is a benchmark for OOP code generation, containing 460 code generation tasks. We selected 230 Java generation tasks from the set for testing. The prompt data used for generation in the experiments was integrated from the information provided by CoderEval. It consolidated all the descriptive information about the target methods available in this project. This facilitated detailed descriptions of the methods and enabled the composition of prompt content for subsequent experimental design.

\subsection{Research FrameWork}

Figure \ref{fig:FrameWork} illustrates the complete process of this experiment, ranging from data preprocessing to code generation and subsequent evaluation. We can use each prompt level for each target method as an example to explain the overall code generation process: First, we extract the corresponding \textbf{Prompt Level} from the \textbf{Dataset}, forming the \textbf{Prompt Data}. Utilizing this data, we invoke the \textbf{OpenAI API} for code generation, obtaining the \textbf{Generated Codes}. Subsequently, all \textbf{Generated codes} undergo unit testing within the corresponding projects in the \textbf{CoderEval Docker}. Finally, we obtain the \textbf{Result Data}, which indicates whether each generated code snippet passed the predefined test cases.

\subsection{Experiment Settings}
We have selected GPT-3.5 (gpt-3.5-turbo-1106) and GPT-4 (gpt-4-1106-preview) as the basic models for our experiments due to their widespread application and superior performance among all LLMs. These models offer exceptional language understanding and generation capabilities, making them highly suitable for complex code generation tasks. Additionally, their extensive use and robust community support ensure the reliability and reproducibility of our experimental results. 

To ensure the comparability of experiments and the consistency of results, all models were configured with uniform parameters. We set the temperature to 0, while keeping the remaining variables at their default configurations. This approach ensures both the equality of experimental conditions and facilitates the subsequent evaluation of each model's baseline performance without specific optimizations.

\subsection{Evaluation Metrics}
\subsubsection{Pass@K}
The Pass@K metric is a crucial indicator for assessing the quality of code generation. It evaluates whether the code generated across multiple attempts passes all test cases at least once. This metric not only prioritizes the correctness of the code but also focuses on the consistency and reliability of the generation system. Moreover, it assesses the capability of code generation tools to successfully generate correct code over multiple attempts. The mathematical principle behind this metric is as follows:

\begin{equation}
\text{pass@}k := \mathbb{E}_{\text{Problems}} \left[ 1 - \frac{{\binom{n-c}{k}}}{{\binom{n}{k}}} \right]
\end{equation}
Where:

\begin{itemize}
    \item $n$ is the total number of samples generated per task.
    \item $c$ is the count of correct samples that pass the unit tests.
    \item $k$ is the parameter of interest in pass@k, representing the number of top attempts considered.
\end{itemize}


\subsubsection{Prompt-Token Cost-Effectiveness}
From an economic perspective, this study innovatively introduces the concept of "Prompt-Token Cost-Effectiveness" (PTCE) to analyze the cost-effectiveness of prompt token information. We first calculated the average token consumption for each prompt level. Then, based on the specific Pass@K and target K value for each prompt level, we could calculate PTCE(K). It is defined as:

\begin{equation}
\text{PTCE(K)} = \frac{\Delta \text{Pass@K}}{\Delta \text{Token Cost}}
\end{equation}
Where:

\begin{flalign}
    &\Delta \text{Pass@K} = \text{Pass@K}_{\text{curr}} - \text{Pass@K}_{\text{prev}}& \\
    &\Delta \text{Token Cost} = \text{Token Cost}_{\text{curr}} - \text{Token Cost}_{\text{prev}}&
\end{flalign}

\subsubsection{Cross-Expert Review}




In software development practice, many metrics do not have an absolute correct solution, particularly those involving subjective evaluations, such as the degree of help to developers. Faced with this complexity, we can draw on the judgments of experienced software development practitioners and experts. Their professional knowledge and practical experience provide us with valuable insights, aiding in more precise classification and assessment of different situations.

In this experiment, to ensure the accuracy of manual evaluations, we hired 10 senior software engineers to participate in the review process. It is important to note that these engineers are not co-authors of this paper. Each engineer has at least six years of experience in object-oriented development, and their professional backgrounds and extensive knowledge provided solid professional support for this study. These practitioners have a wide range of practical experience in the field of software development, ensuring the practicality and reliability of our research results.

To ensure the comprehensiveness and fairness of the evaluations, we employed a cross-review method. Specifically, each generated result was assessed by three independent experts. By comparing the evaluations from all experts, we ensured the consistency of the review outcomes and properly addressed any disagreements that arose. Particularly, for code that received entirely different assessments from three experts, it was reviewed by another three experts in a second round of reviews. If the results of the second review still showed inconsistencies, the most experienced expert served as a mediator, making the final decision based on the existing scores and feedback. This process ensured that our evaluation results were well-substantiated and persuasive.

\subsection{Research Questions}
We have devised a series of experiments to investigate the performance of Large Language Models (LLMs) in various programming tasks, considering multiple key factors: the precision of prompts and its impact on code generation quality, the effect of code task coupling on LLM performance, and the potential of LLMs to alleviate workload in Object-Oriented Programming (OOP) development. Specifically, our research questions are as follows:
\begin{enumerate}
    \item \textbf{RQ1: How does prompt information impact code generation performance?}
    
    Through evaluating the generated code's Pass@K, error type distributions, and cost-effectiveness, we aim to optimize prompt strategies to enhance the practicality of generated code and uncover performance differences between models in the code generation process. This endeavor seeks to deepen our understanding of how prompt richness impacts the introduction of programming errors.
    
    \item \textbf{RQ2: How do prompts assist developers?}
    

    
    We defined different levels of helpfulness value and employed human expert voting to evaluate and analyze the distributions. 

    \item \textbf{RQ3: How does coupling impact generated code?}
    

    Using the concept of Fan-Out, we quantified coupling and further defined three coupling levels. We analyzed the Pass@K for the generated code at each level, thereby specifically examining the influence of the target code's coupling degree on the generation results.
\end{enumerate}



\section{Study Result}
\label{sec:sr}
In this section, we will introduce the main contents of the experiment and discuss the conclusions drawn from these contents. Our study focuses on three research questions (RQs). These RQs include the impact of prompt information richness on code generation, which includes the influence on code generation's Pass@K and the distribution of error types. Additionally, we explore the practical value of code generated by different LLMs and prompt levels in real-world development scenarios. Finally, we discuss the impact of the inherent coupling (Fan-Out) of the target methods on the generated code. These research questions will help us gain a deeper understanding of the potential and challenges of LLMs in code generation.

\subsection{RQ1: How does Prompt Richness Impact Generated Code?}
This study focuses on the impact of prompt information richness on code generation, including Pass@K for generated code and the types of errors encountered. The experiments cover multiple levels of prompt richness, aiming to comprehensively capture how the richness of prompt  affects both the pass rate and the occurrence of errors in the generated code.

\subsubsection{RQ1.1: How does Prompt Richness Impact Pass@K?}
In order to facilitate the measurement of prompt richness, we established a prompt information level table. This encompassed contextual information ranging from the target method to the project level. We divided this information into seven levels, with each level's name, content, and the definition of the changes relative to the previous level, as illustrated in Figure \ref{tab:prompt-content-and-function}.

\begin{table*}[t]
\centering
\small
\caption{Our Definition of Prompts' Levels based on the Information Richness}
\label{tab:prompt-content-and-function}
\begin{tabular}{|c|m{4.5cm}|m{10cm}|}
\hline
\centering \textbf{Level} & \centering \textbf{Prompt Content} \arraybackslash & \centering \textbf{Changes} \arraybackslash \\ \hline
Base-Level\newline(L1) & \texttt{signature} + \texttt{name} & Basic method information: \newline \textbf{signature}: A method's unique identifier, consisting of its name, parameter types, and order. \newline \textbf{name}: Name of the method.\\ \hline
Function-Level \newline (L2-1) & \texttt{signature} + \texttt{name} + \texttt{human\_label} & Compared to L1 added:\newline \textbf{human\_label}: A concise description of the method's functionality. \\ \hline
Doc-Level(L2-2) & \texttt{signature} + \texttt{name} + \texttt{docstring} & Compared to L1 added:\newline \textbf{docstring}: The method's documentation string, providing explanations, functionality, and parameters of the method. \\ \hline
Combo-Level(L3) & \texttt{signature} + \texttt{name} + \texttt{human\_label} + \texttt{docstring} & Compared to L2-1 and L2-2 added:\newline Combined information of \textbf{human\_label} and \textbf{docstring}. \\ \hline
Coupling-Level(L4) & \texttt{signature} + \texttt{name} + \texttt{human\_label} + \texttt{docstring} + \texttt{oracle\_context} & Compared to L3 added:\newline \textbf{oracle\_context}: Provides additional context information about the method, such as used APIs, classes, and variables. \\ \hline
Class-Level(L5) & \texttt{signature} + \texttt{name} + \texttt{human\_label} + \texttt{docstring} + \texttt{oracle\_context} + \texttt{class\_level} & Compared to L4 added:\newline \textbf{class\_level}: Detailed information inside the class, such as class definitions, methods, properties, etc. \\ \hline
Project-Level(L6) & \texttt{signature} + \texttt{name} + \texttt{human\_label} + \texttt{docstring} + \texttt{oracle\_context} + \texttt{class\_level} + \texttt{repo\_level}+ \texttt{level} & Compared to L5 added:\newline \textbf{repo\_level}: Information at the project or code repository level, usually covering a broader context.\newline \textbf{level}: Describes the self-containment of the code snippet, indicating whether the code is independent of other parts. \\ \hline
\end{tabular}
\end{table*}

\begin{figure}[b]
  \centering
  \includegraphics[width=0.9\linewidth]{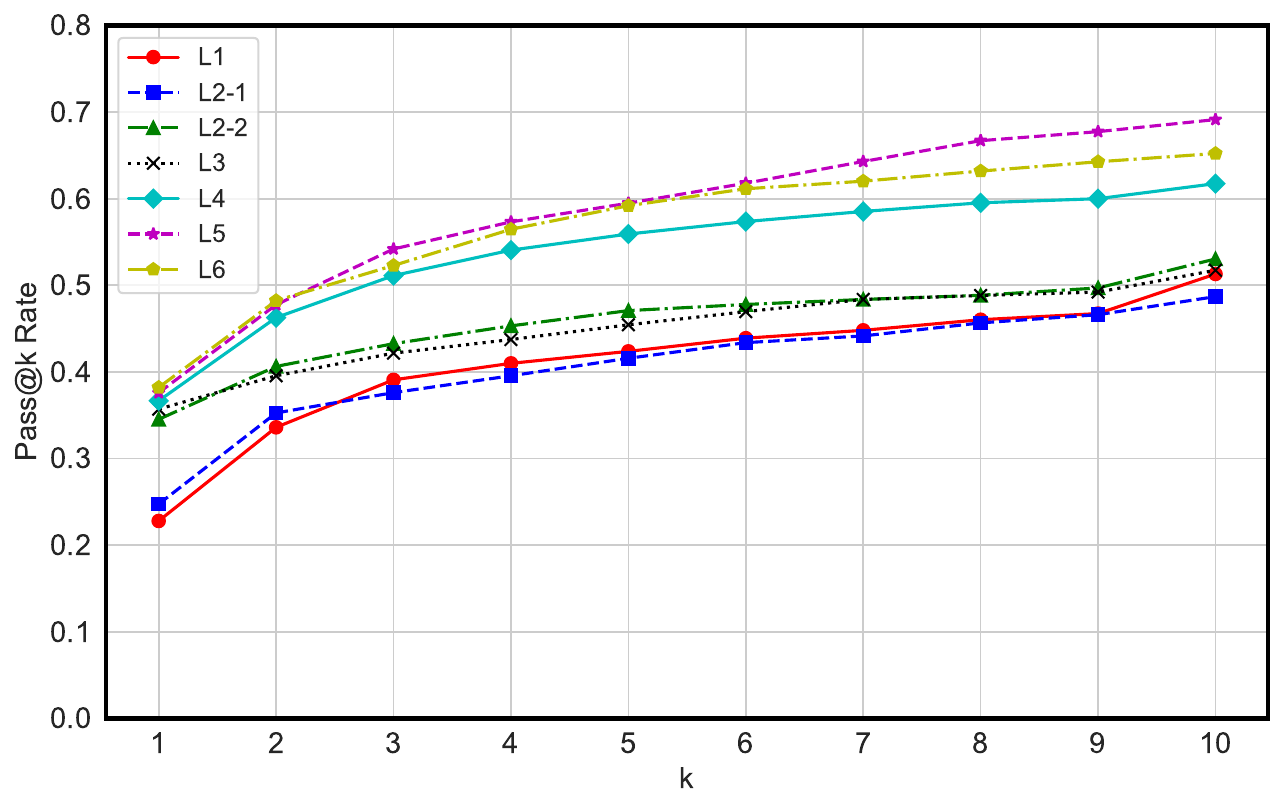}
  \caption{GPT-3.5 Pass@k Performance by Prompt Precision}
  \label{fig:sub1}
\end{figure}

\begin{figure}[b]
  \centering
  \includegraphics[width=0.9\linewidth]{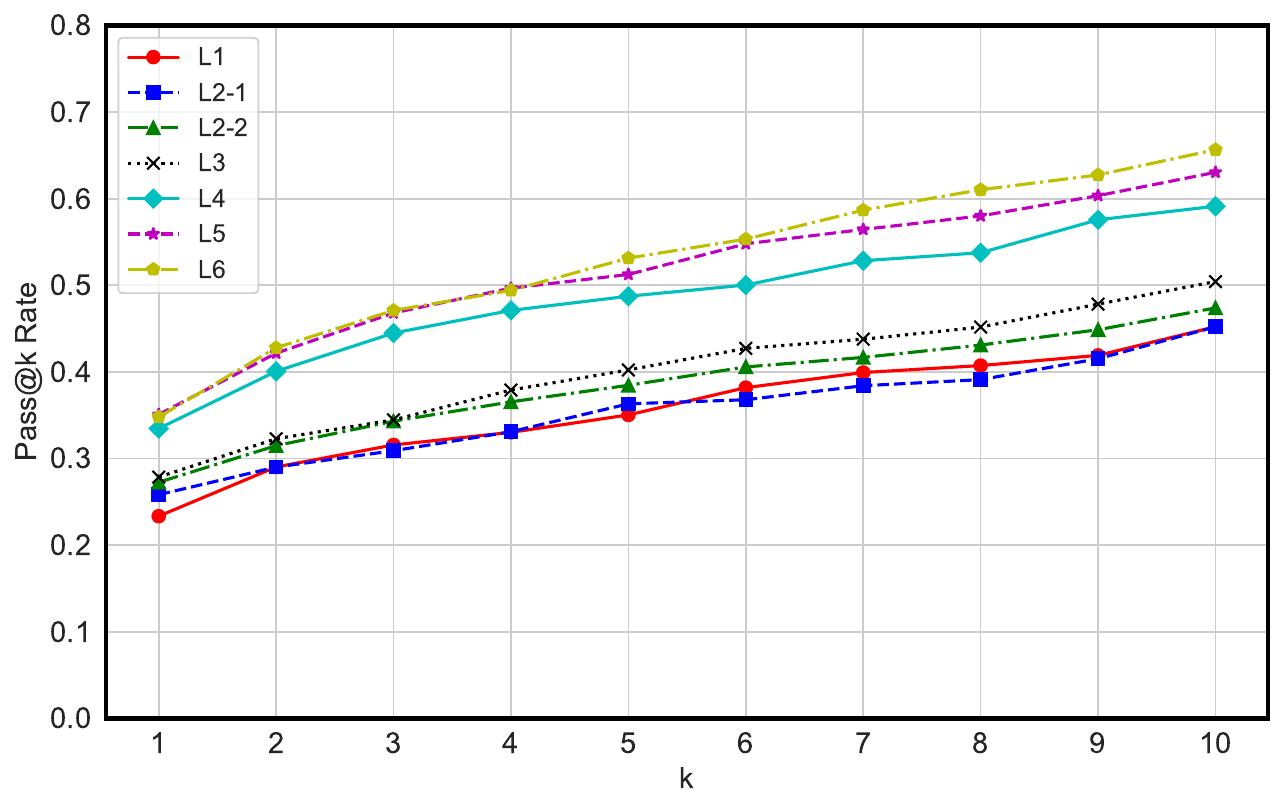}
  \caption{GPT-4 Pass@k Performance by Prompt Precision}
  \label{fig:sub2}
\end{figure}

Figures \ref{fig:sub1} and \ref{fig:sub2} illustrate the changes in code generation Pass@K for GPT-3.5 and GPT-4 across different prompt levels. From the data, it can be clearly observed that as the prompt information increases, the overall Pass@K for generated code exhibits an upward trend. Taking GPT-3.5's Pass@5 as an example, it is 42.4\% at L1, increasing to 55.92\% at L4, and reaching 59.19\% at L6. This demonstrates a significant positive correlation between the richness of prompts and the performance of generated code.  




Further analysis of the figures reveals that the L4 prompt level brought about a significant performance boost. Taking Pass@10 as an example, relative to L3, GPT-3.5 and GPT-4 achieved improvements of 10\% and 8.7\% respectively at L4. Additionally, by comparing the L1 and L2-1 curves across the two models, we found that merely increasing the human-provided humanlabel did not significantly enhance the usability of the code; in some cases (e.g., Pass@7), it even had a negative impact on performance. In contrast, the comparison between L2-1 and L2-2 shows that the document-based signature description outperformed the human description in improving the quality of generated code.  

We also observed some noteworthy anomalies in the figures. Regarding the performance at higher levels (L4 to L6), GPT-4 surprisingly underperformed compared to GPT-3.5, despite being an upgraded, more complex, and larger model based on GPT-3.5. This should be considered an anomaly. Furthermore, although L5 and L6 introduced more context, including class-level context code and project-level invocation information, the corresponding improvement in Pass@K was not significant, and even negative impacts were observed in some cases. This is a crucial consideration for developers, as more information does not necessarily translate to better performance.

\subsubsection{RQ1.2: Cost-Effectiveness Analysis of Prompt Richness}
In practical production scenarios, we also need to consider the cost of utilizing LLMs. This cost is often evaluated based on the number of tokens consumed. In the code generation domain, token consumption largely depends on the richness of the input prompt information. Therefore, this RQ focuses on analyzing the cost-effectiveness of the additional information at each prompt level, termed PTCE. To establish a consistent baseline for calculation, we employ the tiktoken tool provided by OpenAI\footnote{https://github.com/openai/tiktoken} and utilize the cl100k\_base model to compute the token consumption of GPT-3.5 and GPT-4.

\begin{table*}[h]
  \centering
  \caption{Performance Metrics for GPT3.5 and GPT4}
  \label{tab:performance_metrics}
  \setlength{\tabcolsep}{3pt} 
  \begin{tabular}{|c|c|c|c|c|c|c|c|c|c|c|c|c|}
    \hline
    \multicolumn{1}{|c|}{\multirow{2}{*}{\textbf{Metrics}}} & \multicolumn{6}{c|}{GPT3.5} & \multicolumn{6}{c|}{GPT4} \\ \cline{2-13}
    & Pass@3 & PTCE(3) & Pass@5 & PTCE(5) & Pass@10 & PTCE(10) & Pass@3 & PTCE(3) & Pass@5 & PTCE(5) & Pass@10 & PTCE(10) \\
    \hline
    L1 & 39.00\% & - & 42.37\% & - & 51.30\% & - & 31.57\% & - & 35.04\% & - & 45.22\% & - \\
    L2-1 & 37.60\% & -0.0824\% & 41.58\% & -0.0468\% & 48.70\% & -0.1535\% & 30.90\% & -0.0391\% & 36.32\% & 0.0753\% & 45.22\% & 0.0000\% \\
    L2-2 & 43.26\% & 0.1617\% & 47.08\% & 0.1573\% & 53.04\% & 0.1242\% & 34.31\% & 0.0973\% & 38.46\% & 0.0611\% & 47.39\% & 0.0621\% \\
    L3 & 42.17\% & -0.0641\% & 45.43\% & -0.0971\% & 51.74\% & -0.0767\% & 34.46\% & 0.0088\% & 40.23\% & 0.1044\% & 50.43\% & 0.1790\% \\
    L4 & 51.12\% & \textbf{0.2797}\% & 55.92\% & \textbf{0.3278}\% & 61.74\% & \textbf{0.3125}\% & 44.49\% & \textbf{0.3134}\% & 48.74\% & \textbf{0.2659}\% & 59.13\% & \textbf{0.2717}\% \\
    L5 & \textbf{54.10}\% & 0.0055\% & \textbf{59.48}\% & 0.0065\% & \textbf{69.13}\% & 0.0135\% & 46.81\% & 0.0042\% & 51.25\% & 0.0046\% & 63.04\% & 0.0072\% \\
    L6 & 52.29\% & -0.0069\% & 59.18\% & -0.0011\% & 65.22\% & -0.0148\% & \textbf{47.10}\% & 0.0011\% & \textbf{53.14}\% & 0.0071\% & \textbf{65.65}\% & 0.0099\% \\
    \hline
  \end{tabular}
\end{table*}


Table \ref{tab:performance_metrics} presents the token consumption, increment relative to the previous level, and the prompt-token cost-effectiveness at each level of prompt information for K=3, 5, and 10. The data indicates that among all the improvements in prompt information, the transition from L3 to L4 yielded the highest cost-effectiveness. This highlights the importance of incorporating the API, method, and class information directly invoked by the target method (i.e., fan-out information). Furthermore, the process from L4 to L5 incurred the largest token increment due to the addition of context regarding the internal information of the entire class. However, the pass rate improvements brought by L5 and L6 were not significant, resulting in relatively lower cost-effectiveness. This suggests that developers should consider the criticality and economic feasibility of information when designing and implementing prompt strategies, in order to achieve optimal resource utilization and performance enhancement.

\subsubsection{RQ1.3: How does Prompt Richness Affect the Distribution of Error Types?}

\begin{figure}[h]
  \centering
  \includegraphics[width=\linewidth] {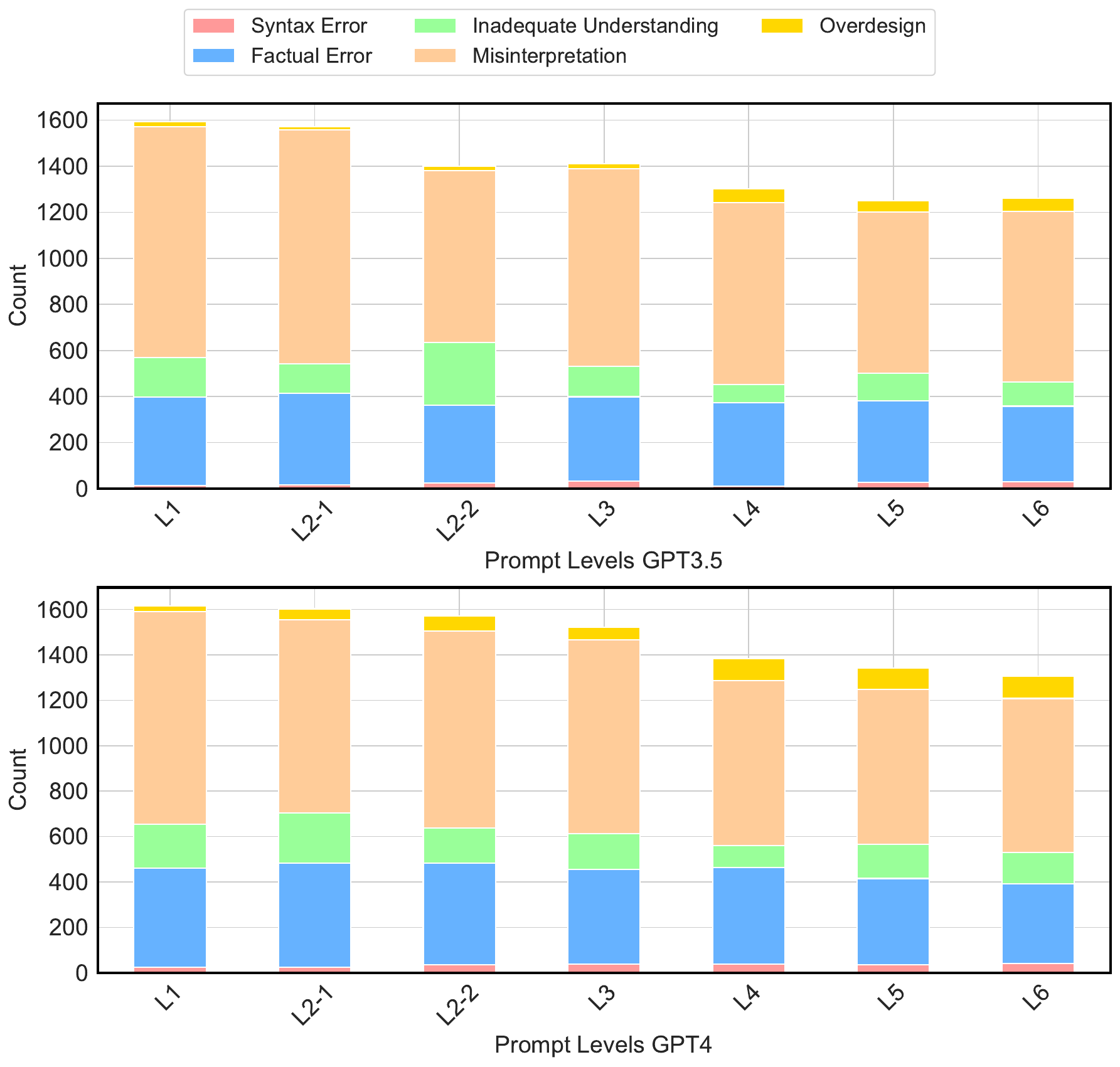}
  \caption{Error Type Distribution by Prompt Level}
  \label{fig:yourlabel}
\end{figure}

In this RQ, we analyzed the error types in the generated code from the perspective of different prompt richness. Specifically, the error types involved in the experiments and their definitions are as follows\cite{huang2023survey}: 

\begin{enumerate}
 \item \textbf{Syntax Error}: Refers to errors in the generated code that do not comply with the rules of the programming language, making it unable to be compiled or interpreted. These errors often involve basic coding format issues, such as mismatched parentheses or missing semicolons.
 \item \textbf{Factual Error}: Refers to errors in the generated code based on incorrect logical assumptions or the use of incorrect programming syntax and features, resulting in discrepancies between the logic or functionality implemented and the expected outcome. For example, using non-existent functions or libraries.
 \item \textbf{Faithfulness Error}: Refers to errors in the generated code that, while syntactically correct and potentially executable, do not meet the user's original requirements or intentions for problem-solving. 
\end{enumerate}
Besides, \textbf{Faithfulness Error} is further categorized as:
  \begin{enumerate}
  \item \textbf{Inadequate Understanding}: The generated code fails to fully implement all expected functionalities, lacking crucial features or functionalities.
  \item \textbf{Misinterpretation}: The generated code does not faithfully reflect the original requirements or intentions, altering the expected behavior.
  \item \textbf{Overdesign}: The generated code introduces unnecessary complexity in attempting to meet requirements, exceeding the original scope of requirements.
  \end{enumerate}
To begin with, we use Tree-Sitter, a tool for detecting syntax questions, to annotate Syntax Error within the erroneous code. Subsequently, for the remaining code data, manual annotation is performed to differentiate between Factual Error and faithfulness errors. For the code annotated as having Faithfulness Error, further annotation is conducted to categorize them into Inadequate Understanding, Misinterpretation, and Overdesign.  The specific distribution of error types is presented in Figure \ref{fig:yourlabel}.

Through the analysis of Figure \ref{fig:yourlabel}, we can clearly observe that as the amount of prompt information gradually increases, the total number of errors in both models shows a decreasing trend, with Misinterpretation errors exhibiting the largest reduction across all error categories. Correspondingly, Syntax Errors and Overdesign issues are on the rise. This indicates that when improving prompt precision, the probability of different error types also changes. Therefore, when utilizing LLMs for code generation, developers need to strike a balance between the quantity and quality of information.

\subsection{RQ2: How do Prompts Help Developers?}
In the process of applying LLMs for code generation in practical development scenarios, encountering erroneous code generation is a common occurrence. Considering that not all erroneous code is entirely valueless, each generated code segment has the potential to assist developers from the perspectives of structure, logic, and other aspects. This experiment aims to evaluate the helpfulness value of the generated code under different levels of prompt information through human evaluation. 

In the experiment, the helpfulness-value of the generated code in assisting developers in writing the target code will be divided into five levels, ranging from 0 to 4, for human evaluation. The definitions of each level are as follows:

\begin{enumerate}
  \item 0: The code segment is completely valueless and irrelevant to the problem.
  \item 1: The code segment has some value, containing information relevant to the problem, but it is easier to write the solution from scratch.
  \item 2: The code segment is somewhat helpful, requiring major changes, but still useful.
  \item 3: The code segment is helpful but has some issues and requires minor modifications to solve the problem.
  \item 4: The code segment is very helpful, highly relevant and informative, and can be directly used to solve the problem.
\end{enumerate}


\begin{figure}[b]
  \centering
  \includegraphics[width=0.8\linewidth]{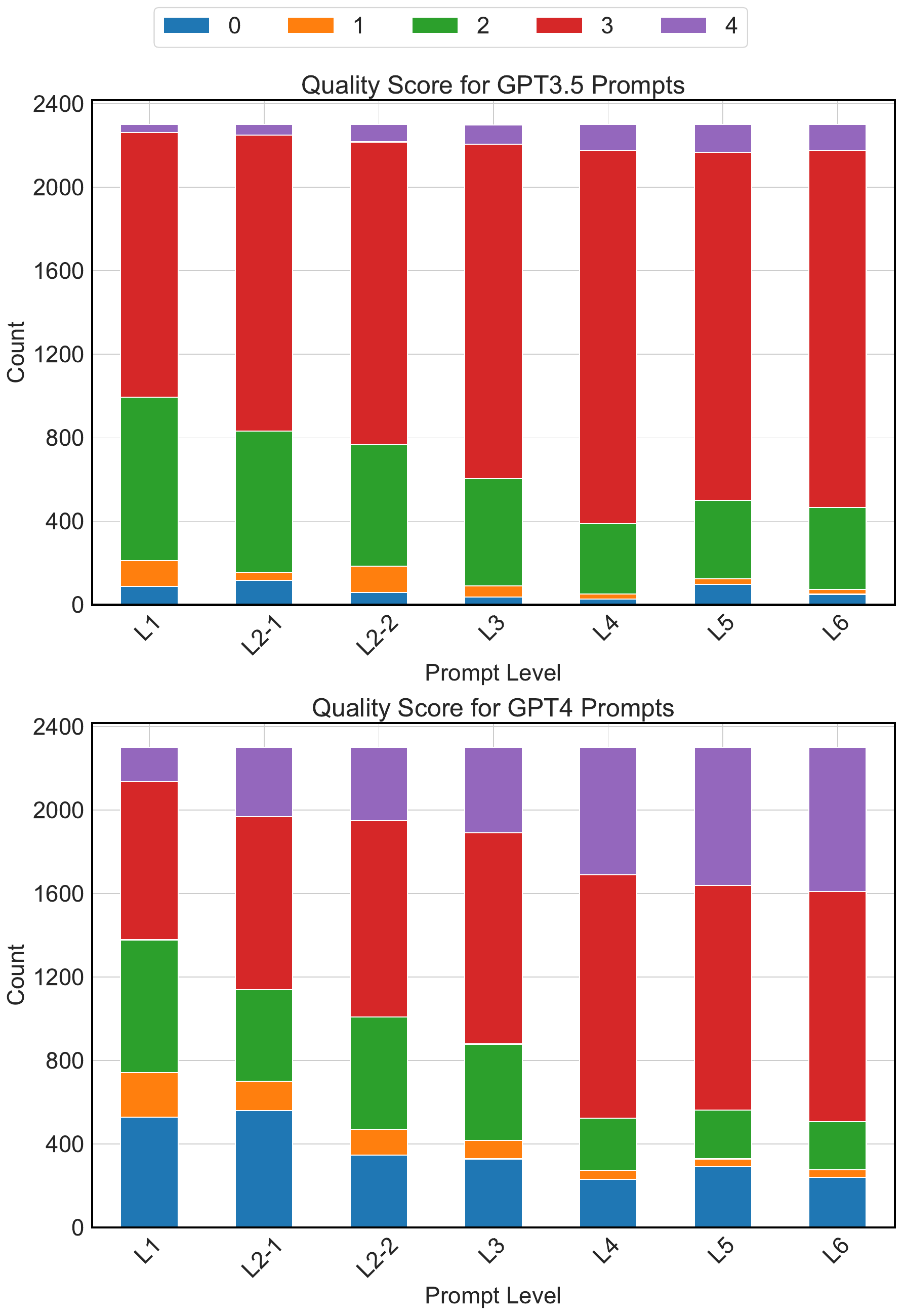}
  \caption{Helpfulness-Value Distribution by Prompt Level}
  \label{fig:Distribution by Prompt Level}
\end{figure}
The annotation results are shown in Figure \ref{fig:Distribution by Prompt Level}. Through observation, we can discern a positive correlation between code helpfulness value and prompt level. As the richness of prompt information increases, the quantity of code with higher helpfulness values gradually rises. Notably, the L4 level exhibits outstanding performance in code generation assistance, with the highest number of codes having a helpfulness value of 3 or above in the distributions of both models. Furthermore, although the overall pass rate of validation is below 50\%, more than half of the codes have a helpfulness value of 2 or higher, indicating that even codes that fail testing possess a certain degree of reference value.\newline

From the perspective of model selection, GPT-4 demonstrates superior performance in generating codes with the highest helpfulness value. In contrast, GPT-3.5 is more prone to producing codes with a helpfulness value of 3 or above across all prompt information levels, suggesting that even at lower prompt levels (such as L1), GPT-3.5 can provide relatively high assistance in practical applications. These observations offer crucial references for developers when choosing models, assisting them in selecting the most suitable model configuration based on specific application requirements to achieve optimal code generation results.

\subsection{RQ3: How does Coupling Impact Generated Code?}

In software engineering, coupling is a metric that measures the degree of interdependence between code modules. To a certain extent, it reflects the complexity and difficulty of writing code. This study aims to investigate the impact of target code coupling on the quality of generated code by LLMs. The experiment is grounded in the concept of fan-out, wherein static code analysis of the target method is performed to extract the sum of the number of APIs, classes, and variables called by each method, serving as a quantitative indicator of coupling.

\begin{figure}[htbp]
  \centering
  \includegraphics[width=0.8\linewidth]{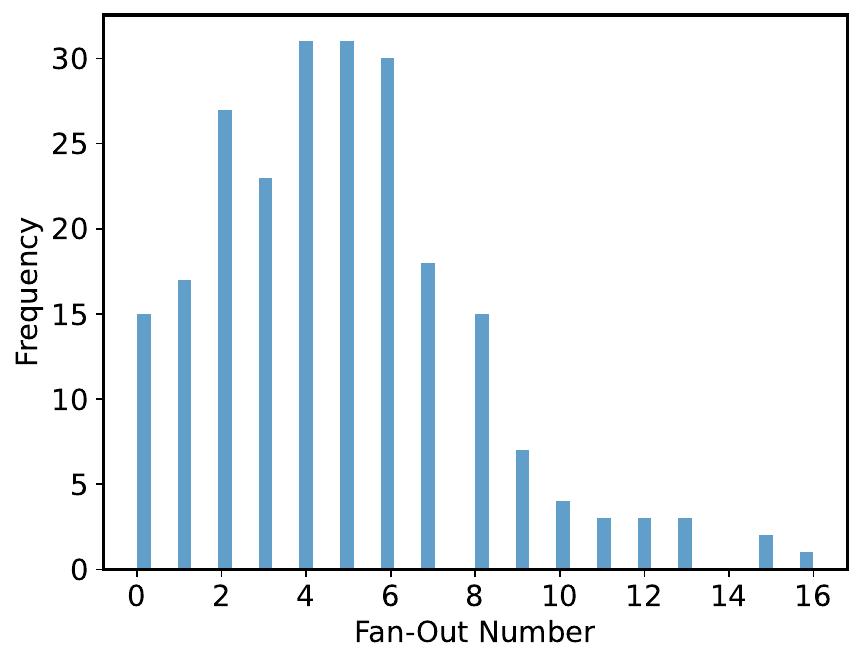}
  \caption{Distribution of Fan-Out Number Of the Dataset}
  \label{fig:Fan-Out}
\end{figure}

The distribution of the quantified measure of coupling is depicted in Figure \ref{fig:Fan-Out}. To enhance the controllability of the analysis, this study categorizes the coupling measure into three classes based on its distribution characteristics: low coupling, medium coupling, and high coupling:
\begin{enumerate}
    \item \textbf{Low Coupling}: Fan-Out <= 3 (covering from the minimum value to near the 25th percentile)
    \item \textbf{Medium Coupling}: 3 < Fan-Out <= 6 (covering a significant portion from the 25th to the 75th percentile)
    \item \textbf{High Coupling}: Fan-Out > 6 (exceeding from the 75th percentile to the maximum value)
\end{enumerate}


Based on this classification, the study will investigate the relationship between coupling and Pass@K.

\begin{figure}[h]
  \centering
  \includegraphics[width=0.9\linewidth]{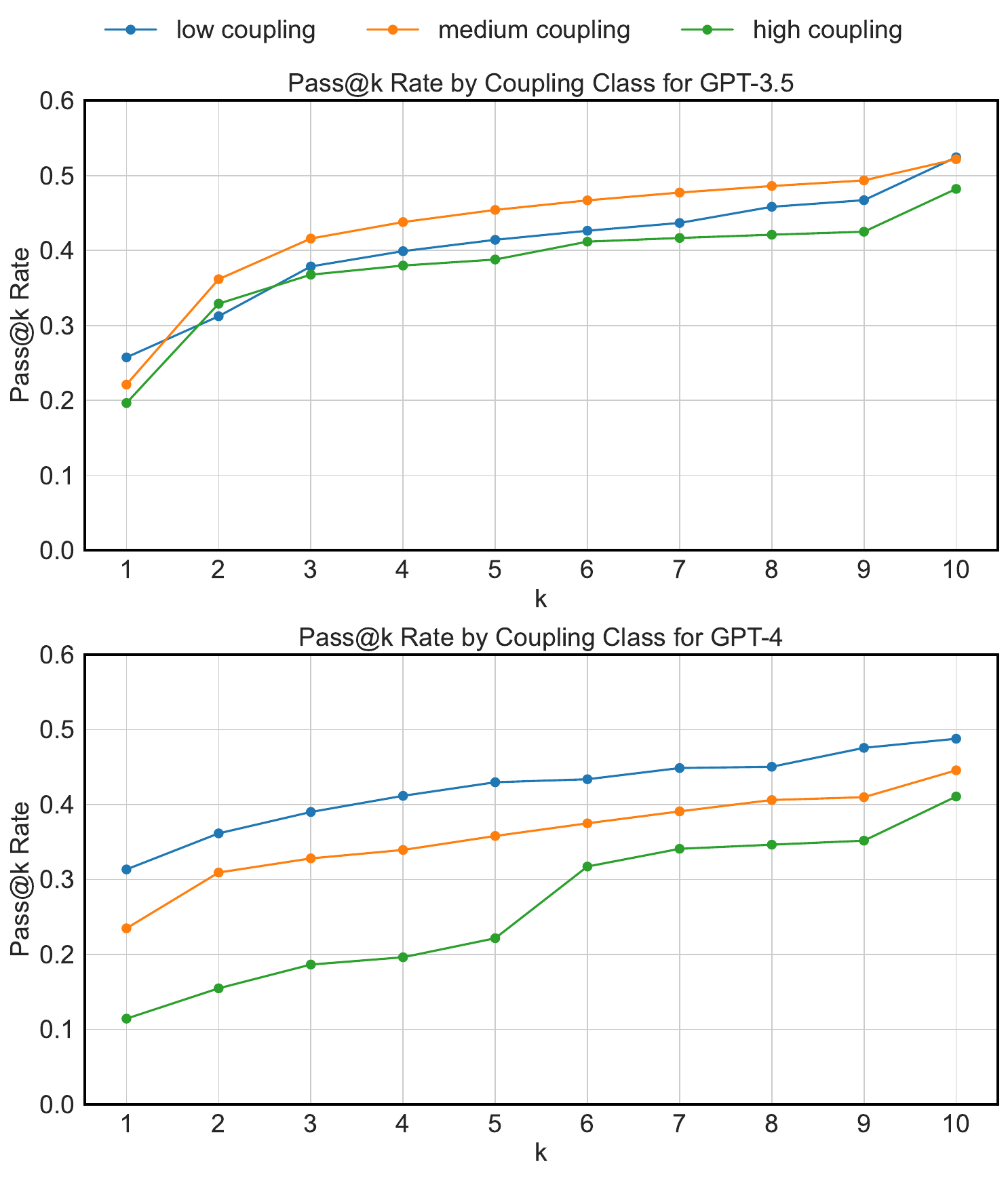}
  \caption{Pass@k Performance by Coupling Class
}
  \label{fig:Coupling}
\end{figure}

As is shown in the Figure \ref{fig:Coupling}, it is evident that  a higher coupling degree of the generation task does not necessarily correspond to a lower Pass@K. This necessitates a comprehensive analysis based on the specific LLM. Different LLMs exhibit varying suitability for different coupling degrees. For instance, GPT-4 performs better on low-coupling tasks but underperforms on high-coupling tasks. Conversely, GPT-3.5 demonstrates superior performance on tasks with medium coupling degrees. Moreover, we note that different models are impacted by coupling degrees to varying extents. Compared to GPT-4, GPT-3.5 is less affected by task coupling degrees, exhibiting more stable Pass@K in code generation. Consequently, developers should consider the actual task circumstances when selecting an LLM. If the task possesses a high coupling degree and requires high stability, GPT-3.5 may be a more suitable choice. However, if the tasks are primarily low-coupling, GPT-4.0 could be a more ideal option.

\section{Discussion}
\label{sec:dis}

In this study, we systematically explore the efficiency and quality of method-level code generation in OOP using large language models such as GPT-3.5 and GPT-4. We discuss the internal and external threats revealed by the experimental content and results, and propose future research directions to address these limitations. Identifying and understanding these potential threats is crucial for ensuring the transparency, reliability, and practical contributions of the research.
\subsection{Threats}
\subsubsection{Internal Threats}
First, \textbf{the dataset used in the experiments has limitations in terms of scale and diversity}. The experiments employed a relatively small dataset from the CoderEval benchmark, comprising only Java tasks, which may limit the generalizability of the results. However, the tasks were carefully chosen from high-quality projects, ensuring the reliability of the findings. Furthermore, the language utilized in the tests is representative of mainstream OOP practices. In future research, larger and more diverse datasets will be incorporated, spanning multiple OOP languages, thereby enhancing the universal applicability of the research outcomes.

Additionally, \textbf{the design suffers from inadequacies in the quantification method for coupling degree}. The assessment of coupling relied primarily on intuitive quantitative accumulation, a simple approach, yet sufficient for providing useful preliminary insights. Future plans include employing static code analysis tools for a more detailed quantitative analysis of coupling.

\subsubsection{External Threats}
In the present experiments, \textbf{the amount of LLMs tested is limited}. The research was primarily confined to GPT-3.5 and GPT-4, limiting the diversity of models, but these have been widely validated as efficient and reliable. Future work will introduce additional models such as Llama3, Codex, PolyCoder, etc., to enhance the comprehensive assessment and comparison of different models' performances.

Furthermore, \textbf{there is a possibility of data leakage in the experiments}. The code data source from CoderEval originates from public GitHub repositories, which may have already been exposed to OpenAI, posing an unavoidable risk. However, the experiment has minimized this impact as much as possible. Future research will employ stricter data validation measures to ensure the accuracy of data research.
These insights not only strengthen the transparency and depth of the research but also provide clear directions for improvement in future studies, aiming to further advance automated programming technologies.

\subsection{Suggestion}
Based on the experimental results, we can offer some practical suggestions to help OOP developers improve efficiency in their actual workflow.

When selecting prompt content, it is advisable to include information about other modules invoked by the target method. This not only helps enhance the quality of code generation but also achieves better cost-effectiveness.

Excessive information may lead to more over-design errors. Developers should not simply assume that more prompt content is better. Instead, the prompt content should be designed within an appropriate scope. This not only avoids excessive token consumption but also prevents wasting developers' time.

When choosing the LLM to use, the coupling degree of the target method needs to be considered. One cannot simply assume that larger LLMs are always better in any situation. For instance, for GPT-4, low-coupling code tasks are more likely to yield better results, while for high-coupling tasks, GPT-3.5 may be more suitable.





With these suggestions, developers can achieve a better code generation experience and work efficiency in practice. By carefully designing prompt content and reasonably selecting LLMs, developers can maximize cost savings while ensuring code quality.

\section{Conclusion}
\label{sec:conc}

This study investigates the influences of prompt richness on the quality of method-level code generation in OOP. Through extensive experiments, we determined that the richness of prompts significantly impacts code quality, aiding developers by maximizing code generation quality without substantial increases in token consumption. Specifically, an optimal level of prompt content, such as L4, offers the best cost-effectiveness. However, our findings also indicate that increasing prompt richness does not uniformly reduce errors; in fact, overly detailed prompts can lead to over-design issues. Furthermore, the usefulness of generated code, as evaluated by experts, suggests that not all incorrect codes are unhelpful. Additionally, LLMs do not always correspond to higher usefulness-value. Our research also highlights the importance of method coupling in code generation: GPT-4 excels in tasks with low coupling, while GPT-3.5 may be better suited for high-coupling scenarios. These insights enable developers to devise more effective prompt strategies and leverage even uncompilable code during development.

In future, this research will expand to involve more diverse LLMs, a broader range of programming languages, and an increased number of testing tasks to enhance the generalizability of our findings. The future work aims to refine automated code generation in OOP by investigating the effects of richer prompt information, potentially using techniques such as vector databases to extend model context. This approach seeks to fine-tune the automated OOP code generation process, thereby improving development efficiency and making significant strides toward more intelligent programming environments.

\balance
\bibliographystyle{ACM-Reference-Format}
\bibliography{bib/references}

\end{document}